\documentclass[11pt,a4paper]{article}

\usepackage{epsfig,amsmath,amssymb,cite}

\begin{document}

\title{Stability of thin-shell wormholes with spherical symmetry}
\author{Ernesto F. Eiroa$^{1,2}$\thanks{e-mail: eiroa@iafe.uba.ar}\\
{\small $^1$ Departamento de F\'{\i}sica, Facultad de Ciencias Exactas y Naturales,} \\ 
{\small Universidad de Buenos Aires, Ciudad Universitaria Pab. I, 1428, 
Buenos Aires, Argentina}\\
{\small $^2$ Instituto de Astronom\'{\i}a y F\'{\i}sica del Espacio, C.C. 67, 
Suc. 28, 1428, Buenos Aires, Argentina}} 

\maketitle

\begin{abstract}
In this article, the stability of a general class of spherically symmetric thin--shell wormholes is studied under perturbations preserving the symmetry. For this purpose, the equation of state at the throat is linearized around the static solutions. The formalism presented here is applied to dilaton wormholes and it is found that there is a smaller range of possible stable configurations for them than in the case of Reissner--Nordstr\"om wormholes with the same charge.\\

\noindent 
PACS number(s): 04.20.Gz, 04.50.+h, 04.40.Nr

\end{abstract}

\section{Introduction}\label{intro}

Traversable Lorentzian wormholes, which have received considerable attention since the paper by Morris and Thorne \cite{motho}, are solutions of the equations of gravitation that connect two regions of the same universe or two separate universes by a throat that allows the passage from one region to the other \cite{motho, visser}. For static geometries, the throat is defined as a minimal area surface that fulfills a flare--out condition \cite{hovis1}. Within the framework of the general relativity theory of gravitation, wormholes must be threaded by exotic matter that violates the null energy condition \cite{motho, visser, hovis1, hovis2}. As it was shown by Visser \textit{et al} \cite{viskardad}, the amount of exotic matter required around the throat can be made as small as needed with a suitable choice of  the geometry of the wormhole.\\

A well known method for constructing wormholes, introduced by Visser \cite{visser, mvis}, is by cutting and pasting two manifolds to form a geodesically complete new one with a shell placed in the joining surface. In this case the exotic matter required for their existence is located at the shell. Stability analysis of thin--shell wormholes under perturbations preserving the original symmetries has been done previously by several authors.  A linearized analysis  of a thin--shell wormhole made by joining two Schwarzschild geometries was carried out by Poisson and Visser \cite{poisson}. This method was applied by Barcelo and Visser to wormholes constructed using branes with negative tensions \cite{barcelo}, and  the case of transparent spherically symmetric thin--shells and wormholes was studied by Ishak and Lake \cite{ishak}. The linearized stability analysis was extended to Reissner--Nordstr\"{o}m thin--shell geometries by Eiroa and Romero \cite{eirom}, and to wormholes with a cosmological constant by Lobo and Crawford \cite{lobo04}. Dynamical thin--shell wormholes were considered by Lobo and Crawford \cite{lobo05}. \\

The presence of matter violating energy conditions is a shared feature of wormholes and modern cosmology, because if general relativity is assumed as the gravity theory describing the large scale behavior of the Universe, the strong energy condition should be violated to explain its accelerated expansion. Models of exotic matter of interest in cosmology have been considered in wormhole construction: phantom energy \cite{phantom}, and a Chaplygin gas \cite{chwh,chaplywh} have been used by several authors as the exotic matter supporting wormholes. Wormholes in higher dimensional spacetimes and in alternative theories of gravitation have been also investigated previously. Studies of wormholes in low energy string theory or in Einstein gravity with a scalar field were done in many articles \cite{string}.  Thin--shell wormholes in dilaton gravity were studied by Eiroa and Simeone \cite{dilwh}, and wormholes in Einstein--Gauss--Bonnet theory were analyzed by several authors \cite{gbwh}. Thin--shell wormholes associated with cosmic strings have been investigated recently \cite{eisi}. Other interesting works can be found in Refs. \cite{other}.\\

In this paper, the stability of a general class of spherically symmetric thin--shell wormholes is analyzed under perturbations preserving the symmetry. With this intention, the equation of state of the exotic matter at the shell is linearized around the static solutions. The analysis is similar to the one previously done for a Chaplygin gas equation of state \cite{chaplywh}. In Sec. \ref{tswh} general thin--shell wormholes are constructed. In Sec. \ref{stab} a general stability formalism is developed, which in Sec. \ref{standard} is shown equivalent to the standard stability method, extended in this work for this purpose. In Sec. \ref{aplic} the stability of thin--shell wormholes with a dilaton field is studied. Finally, in Sec. \ref{conclu} a summary is made and the results are discussed. Units such that $c=G=1$ are used throughout this work.

\section{Thin--shell wormhole construction}\label{tswh}

Let us consider the general spherically symmetric geometry:
\begin{equation} 
ds^2=-f(r)dt^2+f(r)^{-1}dr^2+h(r)(d\theta ^2+\sin^2\theta d\varphi^2),
\label{eq1}
\end{equation}
where $f(r)$ and $h(r)$ are non negative functions from a given value of the radial coordinate. We take two copies of the region with $r\geq a$:
\begin{equation}
\mathcal{M}^{\pm }=\{x/r\geq a\},
\label{eq2a}
\end{equation}
and paste them at the hypersurface
\begin{equation}
\Sigma \equiv \Sigma ^{\pm }=\{x/F(r)=r-a=0\},
\label{eq2b}
\end{equation}
to construct a geodesically complete manifold $\mathcal{M}=\mathcal{M}^{+}\cup \mathcal{M}^{-}$. We choose values of $a$ large enough to avoid the presence of singularities and horizons in the case that the geometry (\ref{eq1}) has any of them. The area of a surface with fixed radius $r$ is $\mathcal{A}=4\pi h(r)$. If $\delta >0$ exists such that $h(r)$ is an increasing function of $r$ for $r\in [a,a+\delta)$, the area $\mathcal{A}$ has a minimum for $r=a$ and the manifold $\mathcal{M}$ represents a wormhole with a throat placed at $\Sigma$. On this manifold we can define a new radial coordinate $l=\pm \int_{a}^{r}g_{rr}dr$ representing the proper radial distance to the throat, which is located at $l=0$; the plus and minus signs correspond, respectively, to $\mathcal{M}^{+}$ and $\mathcal{M}^{-}$. For the analysis we follow the standard Darmois-Israel formalism \cite{daris,mus}. The wormhole throat $\Sigma $ is a synchronous timelike hypersurface, where  we  define coordinates $\xi ^{i}=(\tau ,\theta,\varphi )$, with $\tau $\ the proper time on the shell. For the stability analysis under perturbations preserving the symmetry we let the radius of the throat be a function of time: $a(\tau)$. We assume that the geometry remains static outside the throat, regardless that the throat radius can vary with time, so no gravitational waves are present. This is guaranteed if a Birkhoff theorem holds for the original metric\footnote{See Refs. \cite{brokome} for the conditions that should be satisfied in a spherically symmetric spacetime to have a Birkhoff theorem.}. The second fundamental forms (extrinsic curvature) associated with the two sides of the shell are:
\begin{equation}
K_{ij}^{\pm }=-n_{\gamma }^{\pm }\left. \left( \frac{\partial ^{2}X^{\gamma
} } {\partial \xi ^{i}\partial \xi ^{j}}+\Gamma _{\alpha \beta }^{\gamma }
\frac{ \partial X^{\alpha }}{\partial \xi ^{i}}\frac{\partial X^{\beta }}{
\partial \xi ^{j}}\right) \right| _{\Sigma },
\label{eq3}
\end{equation}
where $n_{\gamma }^{\pm }$ are the unit normals ($n^{\gamma }n_{\gamma }=1$) 
to $\Sigma $ in $\mathcal{M}$:
\begin{equation}
n_{\gamma }^{\pm }=\pm \left| g^{\alpha \beta }\frac{\partial F}{\partial
X^{\alpha }}\frac{\partial F}{\partial X^{\beta }}\right| ^{-1/2}
\frac{\partial F}{\partial X^{\gamma }}.
\label{eq4}
\end{equation}
Adopting the orthonormal basis $\{e_{\hat{\tau}},e_{\hat{\theta}},
e_{\hat{\varphi}}\}$ ($e_{\hat{\tau}}=e_{\tau }$, $e_{\hat{\theta}}=[h(a)]^{-1/2}e_{\theta }$, $e_{\hat{\varphi}}=[h(a)\sin^{2} \theta ]^{-1/2}e_{\varphi }$), for the metric (\ref{eq1}), we obtain that
\begin{equation}
K_{\hat{\theta}\hat{\theta}}^{\pm }=K_{\hat{\varphi}\hat{\varphi}}^{\pm
}=\pm \frac{h'(a)}{2 h(a)}\sqrt{f(a)+\dot{a}^2},
\label{eq5}
\end{equation}
and
\begin{equation}
K_{\hat{\tau}\hat{\tau}}^{\pm }=\mp \frac{2\ddot{a}+f'(a)}{2\sqrt{f(a)+\dot{a}^2}} ,
\label{eq6}
\end{equation}
where the prime and the dot represent the derivatives with respect to $r$ and $\tau$, respectively. With the definitions of $[K_{_{\hat{\imath}\hat{\jmath}}}]\equiv K_{_{\hat{\imath}\hat{\jmath}}}^{+}-K_{_{\hat{\imath}\hat{\jmath}}}^{-}$, and  $K=tr[K_{\hat{\imath}\hat{\jmath }}]=[K_{\; \hat{\imath}}^{\hat{\imath}}]$, and the introduction of the surface stress-energy tensor $S_{_{\hat{\imath}\hat{
\jmath} }}={\rm diag}(\sigma ,p_{\hat{\theta}},p_{\hat{\varphi}})$ we have the Einstein equations on the shell (also called the Lanczos equations):
\begin{equation}
-[K_{\hat{\imath}\hat{\jmath}}]+Kg_{\hat{\imath}\hat{\jmath}}=8\pi 
S_{\hat{\imath}\hat{\jmath}},
\label{eq7}
\end{equation}
that in our case results in a shell of radius $a$ with energy density 
$\sigma$ and transverse pressure $p=p_{\hat{\theta}}=p_{\hat{\varphi}}$ 
given by
\begin{equation}
\sigma=-\frac{\sqrt{f(a)+\dot{a}^2}}{4\pi}\frac{h'(a)}{h(a)},
\label{eq8a}
\end{equation}
\begin{equation}
p=\frac{\sqrt{f(a)+\dot{a}^2}}{8\pi}\left[ \frac{2\ddot{a}+f'(a)}{f(a)+\dot{a}^2}+\frac{h'(a)}{h(a)}\right] .
\label{eq8b}
\end{equation}
The flare-out condition, i.e. that the area is minimal at the throat (then $h(r)$ increases for $r$ close to $a$ and $h'(a)>0$), implies that the energy density is negative at the throat, so we have exotic matter there. The values of energy density and pressure corresponding to static wormholes are obtained by putting the time derivatives equal to zero in the equations above.

\section{Stability analysis: formalism}\label{stab}

The static solutions with throat radius $a_{0}$ have energy density and pressure at the throat given by
\begin{equation} 
\sigma_{0}=-\frac{\sqrt{f(a_{0})}}{4\pi}\frac{h'(a_{0})}{h(a_{0})},
\label{eq9a}
\end{equation}
and
\begin{equation} 
p_{0}=\frac{\sqrt{f(a_{0})}}{8\pi}\left[ \frac{f'(a_{0})}{f(a_{0})}+\frac{h'(a_{0})}{h(a_{0})}\right] .
\label{eq9b}
\end{equation}
To study the stability of these static solutions under perturbations preserving the spherical symmetry we linearize the equation of state around the static solution:
\begin{equation}
p-p_{0}=\beta_{0}^2 (\sigma -\sigma_{0}),
\label{eq10}
\end{equation} 
where $\beta_{0}^2 = (\partial p / \partial \sigma ) | _{\sigma_0}$ is a parameter. For ordinary matter, $\beta _{0}$ represents the velocity of sound, so it should satisfy $0< \beta _{0}^{2}\leq 1$. If the matter is exotic, as it happens to be in the throat,  $\beta _{0}$ is not necessarily the velocity of sound and it is not clear which values it can take (see discussion in Ref. \cite{poisson}). Then, if we replace Eqs. (\ref{eq8a}-\ref{eq9b}) in Eq. (\ref{eq10}), we have the differential equation to be satisfied by the throat radius:
\begin{equation}
\ddot{a}+(1+2\beta_{0}^2)\frac{h'(a)}{2h(a)}(f(a)+\dot{a}^2)+\frac{f'(a)}{2}-\left[(1+2\beta_{0}^2)\frac{h'(a_{0})}{2h(a_{0})}f(a_{0})+\frac{f'(a_{0})}{2}\right] \frac{\sqrt{f(a)+\dot{a}^2}}{\sqrt{f(a_{0})}}=0.
\label{eq11}
\end{equation}
Close to $a_{0}$ we can rewrite $a(\tau )$ in the form
\begin{equation}
a(\tau )=a_{0}[1+\epsilon (\tau )],
\label{eq12} 
\end{equation} 
with $|\epsilon (\tau )|\ll 1$ a small radial perturbation. By replacing Eq. (\ref{eq12}) in Eq. (\ref{eq11}) and defining $\nu (\tau)= \dot{\epsilon }(\tau)$, Eq. (\ref{eq11}) can be written as a set of first order differential equations
\begin{equation}
\left\{
\begin{array}{l}
\dot{\epsilon}=\nu \\
\dot{\nu }=\alpha(\epsilon,\nu)
\end{array}
\right.
\label{eq13}
\end{equation} 
with
\begin{eqnarray}
\alpha(\epsilon,\nu) & =& \left[(1+2\beta_{0}^2)\frac{h'(a_{0})}{2h(a_{0})}f(a_{0})+\frac{f'(a_{0})}{2}\right] \frac{\sqrt{f[a_{0}(1+\epsilon)]+a_{0}^2\nu^2}}{a_{0}\sqrt{f(a_{0})}} \nonumber \\
& & -(1+2\beta_{0}^2)\frac{h'[a_{0}(1+\epsilon)]}{2a_{0}h[a_{0}(1+\epsilon)]}\left\{f[a_{0}(1+\epsilon)]+a_{0}^2\nu^2\right\}-\frac{f'[a_{0}(1+\epsilon)]}{2a_{0}}.
\label{eq14}
\end{eqnarray} 
Taylor expanding to first order in $\epsilon $ and $\nu $ we obtain
\begin{equation}
\dot{\xi}=M\xi,
\label{eq15} 
\end{equation} 
where
\begin{equation}
\xi=\left(\begin{array}{c}
\epsilon \\
\nu
\end{array}\right),
\;\; \; \;\; 
M=\left(\begin{array}{cc}
0 & 1 \\ 
\Delta & 0
\end{array}\right)
\label{eq16}
\end{equation} 
and
\begin{equation}
\Delta=\frac{[f'(a_{0})]^2}{4f(a_{0})}-\frac{f''(a_{0})}{2}+(1+2\beta_{0}^2)\frac{2f(a_{0})[h'(a_{0})]^2-2f(a_{0})h(a_{0})h''(a_{0})-f'(a_{0})h(a_{0})h'(a_{0})}{4[h(a_{0})]^2}.
\label{eq17} 
\end{equation} 
When $\Delta>0$ the matrix $M$ has two real eigenvalues: $\lambda_{1}=-\sqrt{\Delta}<0$ and $\lambda_{2}=\sqrt{\Delta}>0$. This case is unstable because of the presence of an eigenvalue with positive real part. The imaginary parts of the eigenvalues are zero, so the instability is of saddle type. If $\Delta=0$, both eigenvalues are zero: $\lambda_{1}=\lambda_{2}=0$, and to first order in $\epsilon$ and $\nu$ we obtain $\nu=\mathrm{constant}=\nu_{0}$ and $\epsilon=\epsilon_{0}+\nu_{0}(\tau-\tau_{0})$, then the static solution is unstable. When $\Delta<0$ there are two imaginary eigenvalues $\lambda_{1}=-i\sqrt{|\Delta |}$ and $\lambda_{2}=i\sqrt{|\Delta |}$; then the linear system does not determine the stability and the set of nonlinear differential equations should be taken into account. In this case, it is useful to rewrite Eq. (\ref{eq13}) in polar coordinates $(\rho, \gamma )$, with $\epsilon=\rho \cos \gamma $ and $\nu=\rho \sin \gamma $, and make a first order Taylor expansion in $\rho$, which gives
\begin{equation}
\left\{
\begin{array}{l}
\dot{\rho}=\sin \gamma \cos \gamma (1+\Delta)\rho \\
\dot{\gamma}=\Delta\cos^2\gamma -\sin^2\gamma +\omega(\gamma )\rho ,
\end{array}
\right.
\label{eq18}
\end{equation}
where $\omega(\gamma )$ is a bounded periodic function of $\gamma $. Close to the equilibrium point, i.e. for small values of $\rho$, the time derivative of the angle $\gamma $ is negative, because the leading term $\Delta \cos^2\gamma -\sin^2 \gamma <0$, so $\gamma $ is a monotonous decreasing function of time, and then the solution curves rotate clockwise around the equilibrium point. To see that these solution curves are closed orbits for small $\rho$, we can take a time $\tau_{1}$ so that $(\epsilon(\tau_{1}),\nu(\tau_{1}))=(\epsilon_{1},0)$ with $\epsilon_{1}>0$. As the solution curve passing through $(\epsilon_{1},0)$ rotates clockwise around $(0,0)$, there will be a time $\tau_{2}>\tau_{1}$ such that the curve will cross the $\epsilon$ axis again, in the point $(\epsilon(\tau_{2}),\nu(\tau_{2}))=(\epsilon_{2},0)$, with $\epsilon_{2}<0$. It is easy to see that Eq. (\ref{eq13}) is invariant under the transformation composed of a time inversion $\tau\rightarrow-\tau$ and the inversion $\nu\rightarrow-\nu$, so the counterclockwise curve beginning in $(\epsilon_{1},0)$ should cross the $\epsilon$ axis also in $(\epsilon_{2},0)$. Therefore, for $\Delta<0$ the solution curves of Eq. (\ref{eq13}) should be closed orbits near the equilibrium point $(0,0)$, which is a stable center. The only stable static solutions with throat radius $a_{0}$ are then those which have $\Delta<0$, and they are not asymptotically stable, i.e. when perturbed the throat radius oscillates periodically around the equilibrium radius, without settling down again.

\section{The standard approach}\label{standard}

In this Section, the standard stability method for thin--shell wormholes, based on the definition of a potential \cite{poisson,ishak,eirom,lobo04,lobo05}, is extended to any metric of the form (\ref{eq1}), and it is compared with the formalism developed in the previous Section. From the Eqs. (\ref{eq8a}) and (\ref{eq8b}), it is easy to verify the energy conservation equation:
\begin{equation}
\frac{d}{d\tau }\left( \sigma \mathcal{A}\right) +p\frac{d\mathcal{A}}{d\tau }=
\left\lbrace \left[ h'(a)\right]^2 -2h(a)h''(a)\right\rbrace \frac{\dot{a}\sqrt{f(a)+ \dot{a}^2}}{2h(a)},
\label{p1}
\end{equation}
where $\mathcal{A}=4\pi h(a)$ is the area of the wormhole throat. The first term in the left hand side of Eq. (\ref{p1}) represents the internal energy change of the throat and the second the work done by the internal forces of the throat, while the term in the right hand side represents a flux. Eq. (\ref{p1})  can be written in the form
\begin{equation}
h(a)\dot{\sigma}+h'(a)\dot{a}(\sigma +p)=-\left\lbrace \left[ h'(a)\right]^2 -2h(a)h''(a)\right\rbrace
\frac{\dot{a}\sigma }{2 h'(a)}.
\label{p2}
\end{equation}
When $[h'(a)]^2 -2h(a)h''(a)=0$, the flux term in the right hand side of Eqs. (\ref{p1}) and (\ref{p2}) is zero\footnote{This is equivalent to the denominated transparency condition \cite{ishak,lobo05}.} and they take the form of simple conservation equations. This happens if $h(a)=C(a+D)^2$, with $C>0$ and $D$ constants or if $h(a)=C$; the second case being unphysical, since there is no throat. Using that $\sigma ^{\prime}=\dot{\sigma }/\dot{a}$, from Eq. (\ref{p2}) one obtains:
\begin{equation}
h(a)\sigma ^{\prime}+h'(a)(\sigma +p)+\left\lbrace \left[ h'(a)\right]^2 -2h(a)h''(a)\right\rbrace
\frac{\sigma }{2 h'(a)}=0,
\label{p3}
\end{equation}
which, if $p$ is known as a function of $\sigma $, is a first order differential that can be integrated\footnote{Eq. (\ref{p3}) can be recast in the form $\sigma ^{\prime}(a)=\mathcal{F}(a, \sigma (a))$, for which always exists a unique solution with a given initial condition, provided that $\mathcal{F}$ has continuous partial derivatives.} to obtain $\sigma (a)$. Thus, replacing $\sigma (a)$ in Eq. (\ref{eq8a}) and regrouping terms, the dynamics of the wormhole throat is completely determined by a single equation:
\begin{equation}
\dot{a}^{2}=-V(a),
\label{p4}
\end{equation}
with 
\begin{equation}
V(a)=f(a)-16\pi ^{2}\left[ \frac{h(a)}{h'(a)}\sigma (a)\right] ^{2}.
\label{p5}
\end{equation}
A Taylor expansion to second order of the potential $V(a)$ around the static solution yields:
\begin{equation}
V(a)=V(a_{0})+V^{\prime }(a_{0})(a-a_{0})+\frac{V^{\prime \prime }(a_{0})}{2}
(a-a_{0})^{2}+O(a-a_{0})^{3}.
\label{p6}
\end{equation}
From Eq. (\ref{p5}) the first derivative of $V(a)$ is
\begin{equation}
V^{\prime }(a)=f^{\prime }(a)-32\pi ^{2}\sigma (a) \frac{h(a)}{h'(a)} \left\lbrace  \left[ 1-\frac{h(a)h''(a)}{[h'(a)]^2}\right] \sigma (a) +\frac{h(a)}{h'(a)}\sigma ^{\prime }(a)\right\rbrace ,
\label{p7}
\end{equation}
which using Eq. (\ref{p3}) takes the form
\begin{equation}
V^{\prime }(a)=f^{\prime }(a)+16\pi ^{2}\sigma (a)\frac{h(a)}{h'(a)} \left[ \sigma (a)+2p(a)\right]  .
\label{p8}
\end{equation}
The second derivative of the potential is
\begin{eqnarray}
V^{\prime \prime }(a)&=&f^{\prime \prime }(a)+16\pi ^{2}\left\lbrace \left[ \frac{h(a)}{h'(a)}\sigma ^{\prime }(a)+\left( 1-\frac{h(a)h''(a)}{[h'(a)]^2} \right)  \sigma (a)\right] \left[ \sigma (a)+2p(a)\right] \right.
\\ \nonumber
& & \left. +\frac{h(a)}{h'(a)}\sigma (a) \left[ \sigma ^{\prime }(a)+2p^{\prime }(a)\right] \right\rbrace  .
\label{p9}
\end{eqnarray}
Since $\sigma ^{\prime }(a)+2p^{\prime }(a)=\sigma ^{\prime }(a)[1+2p^{\prime }(a)/\sigma ^{\prime }(a)]$, replacing the parameter $\beta ^{2}= dp/d\sigma =p^{\prime }/\sigma ^{\prime }$, we have that $\sigma ^{\prime }(a)+2p^{\prime }(a)=\sigma ^{\prime }(a)(1+2\beta ^{2})$, and using Eq. (\ref{p3}) again, we obtain 
\begin{equation}
V^{\prime \prime }(a)=f^{\prime \prime }(a)-8\pi ^{2}\left\lbrace \left[ \sigma (a)+2p(a)\right] ^{2}+2\sigma (a)\left[ \left(  \frac{3}{2}-\frac{h(a)h''(a)}{[h'(a)]^2}\right)\sigma (a)+p(a)\right] (1+2\beta ^{2})\right\rbrace .
\label{p10}
\end{equation}
Using Eqs. (\ref{eq9a}) and (\ref{eq9b}), it is not difficult to see that $V(a_{0})=V^{\prime }(a_{0})=0$, so the potential is
\begin{equation}
V(a)=\frac{1}{2}V^{\prime \prime }(a_{0})(a-a_{0})^{2}+O[(a-a_{0})^{3}],
\label{p11}
\end{equation}
with 
\begin{equation}
V^{\prime \prime }(a_{0})= f^{\prime \prime }(a_{0})-\frac{[f^{\prime }(a_{0})]^2}{2f(a_{0})}-(1+2\beta _{0}^{2})\frac{2f(a_{0})[h'(a_{0})]^2-2f(a_{0})h(a_{0})h''(a_{0})-f'(a_{0})h(a_{0})h'(a_{0})}{2[h(a_{0})]^2},
\label{p12}
\end{equation}
and $\beta _{0}=\beta (\sigma _{0})$. The wormhole is stable if and only if $V^{\prime \prime }(a_{0})>0$. By comparing Eqs. (\ref{eq17}) and (\ref{p12}) it is easy to check that $V^{\prime \prime }(a_0)=-2\Delta $. As it was shown in Sec. \ref{stab}, the wormholes are stable only when $\Delta <0$, so both methods are equivalent.

\section{Wormholes with a dilaton field}\label{aplic}

In this section we apply the formalism developed in Sec. \ref{stab} to dilaton thin--shell wormholes. For comparison, we first recover the results corresponding to Schwarzschild and Reissner--Norsdtr\"om wormholes, previously studied in the literature. The metric functions corresponding to the Reissner--Nordstr\"{o}m geometry are:
\begin{equation}
f(r)  = 1-\frac{2M}{r}+\frac{Q^2}{r^2}, \;\;\;\;\;
h(r)  = r^2,
\label{rn0}
\end{equation}
where $M$ is the mass and $Q$ is the charge. Replacing these functions in the Eqs. (\ref{eq9a}) and (\ref{eq9b}) we recover the energy density and pressure corresponding to the thin-shell wormhole associated with the Reissner--Nordstr\"{o}m geometry, obtained in Ref. \cite{eirom}:
\begin{equation}
\sigma _{0}=-\frac{1}{2\pi a_{0}}\sqrt{1-\frac{2M}{a_{0}}+\frac{Q^{2}}{a_{0}^{2}}},
\label{rn1}
\end{equation}
\begin{equation}
p_{0}=\frac{1}{4\pi a_{0}}\frac{1-\frac{M}{a_{0}}}{\sqrt{1-\frac{2M}{a_{0}}+
\frac{Q^{2}}{a_{0}^{2}}}},
\label{rn2}
\end{equation}
where the allowed values of the throat radius are $a_{0}>r_{h}=M+\sqrt{M^2-Q^2}$ if $|Q|\le M$ and  $a_{0}>0$ if $|Q|>M$. From Eq. (\ref{eq17}) $\Delta $ has the form
\begin{equation}
\Delta=\frac{1}{a_{0}^{4}}\left[ \frac{a_{0}\left[ 
(a_{0}-M)^{3}+M(M^{2}-Q^{2})\right] }{a_{0}^{2}-2Ma_{0}+Q^{2}}
+2\left( a_{0}^{2}-3Ma_{0}+2Q^{2}\right) \beta _{0}^{2}\right],
\label{rn3}
\end{equation}
which, from the condition $\Delta <0$ and defining the auxiliary function
\begin{equation}
\chi(a_{0})\equiv \frac{-a_{0}\left[ (a_{0}-M)^{3}+
M(M^{2}-Q^{2})\right] }{2\left( a_{0}^{2}-2Ma_{0}+Q^{2}\right) \left(
a_{0}^{2}-3Ma_{0} +2Q^{2}\right) },
\nonumber
\end{equation} 
gives five possible cases accordingly to the value of charge:
\begin{enumerate}
\item  Case $0\leq \frac{|Q|}{M}<1$. Stable when\\
i) $\beta _{0}^{2}>\chi(a_{0})$ if $1+\sqrt{1-\frac{Q^{2}}{M^{2}}}<\frac{
a_{0}}{M}<\frac{3}{2}+\frac{1}{2}\sqrt{9-8\frac{Q^{2}}{M^{2}}}$, \\
or\\
ii) $\beta _{0}^{2}<\chi(a_{0})$ if $\frac{a_{0}}{M}>\frac{3}{2}+\frac{1}{2}
\sqrt{9-8\frac{Q^{2}}{M^{2}}}$.
\item  Case $\frac{|Q|}{M}=1$. Stable when\\
i) $\beta _{0}^{2}>\chi(a_{0})$ if $1<\frac{a_{0}}{M}<2$, \\
or\\
ii) $\beta _{0}^{2}<\chi(a_{0})$ if $\frac{a_{0}}{M}>2$.
\item  Case $1<\frac{|Q|}{M}<\frac{3}{\sqrt{8}}$. Stable when\\
i) $\left\{ 
\begin{array}{ll}
\beta _{0}^{2}<\chi(a_{0}) & 
\mbox{if $0< \frac{a_{0}}{M}<\frac{3}{2}-\frac{1}{2}
\sqrt{9-8\frac{Q^{2}}{M^{2}}}$} \\ 
\beta _{0}^{2}\in \mathbb{R} & 
\mbox{if $\frac{a_{0}}{M}=\frac{3}{2}-\frac{1}{2}
\sqrt{9-8\frac{Q^{2}}{M^{2}}}$} \\ 
\beta _{0}^{2}>\chi(a_{0}) & 
\mbox{if $\frac{3}{2}-\frac{1}{2}
\sqrt{9-8\frac{Q^{2}}{M^{2}}}<\frac{a_{0}}{M}<\frac{3}{2}+\frac{1}{2}
\sqrt{9-8\frac{Q^{2}}{M^{2}}}$}
\end{array}
\right. $\\
or\\
ii) $\beta _{0}^{2}<\chi(a_{0})$ if $\frac{a_{0}}{M}>\frac{3}{2}+\frac{1}{2}
\sqrt{9-8\frac{Q^{2}}{M^{2}}}$.
\item  Case $\frac{|Q|}{M}=\frac{3}{\sqrt{8}}$. Stable when\\
i) $\beta _{0}^{2}<\chi(a_{0})$ if $0<\frac{a_{0}}{M}<\frac{3}{2}$, \\
or\\
ii) $\beta _{0}^{2}<\chi(a_{0})$ if $\frac{a_{0}}{M}>\frac{3}{2}$.
\item  Case $\frac{|Q|}{M}>\frac{3}{\sqrt{8}}$. Stable when\\
$\beta _{0}^{2}<\chi(a_{0})$ if $\frac{a_{0}}{M}>0$.
\end{enumerate}
\begin{figure}[t!]
\begin{center}
\includegraphics[width=17cm]{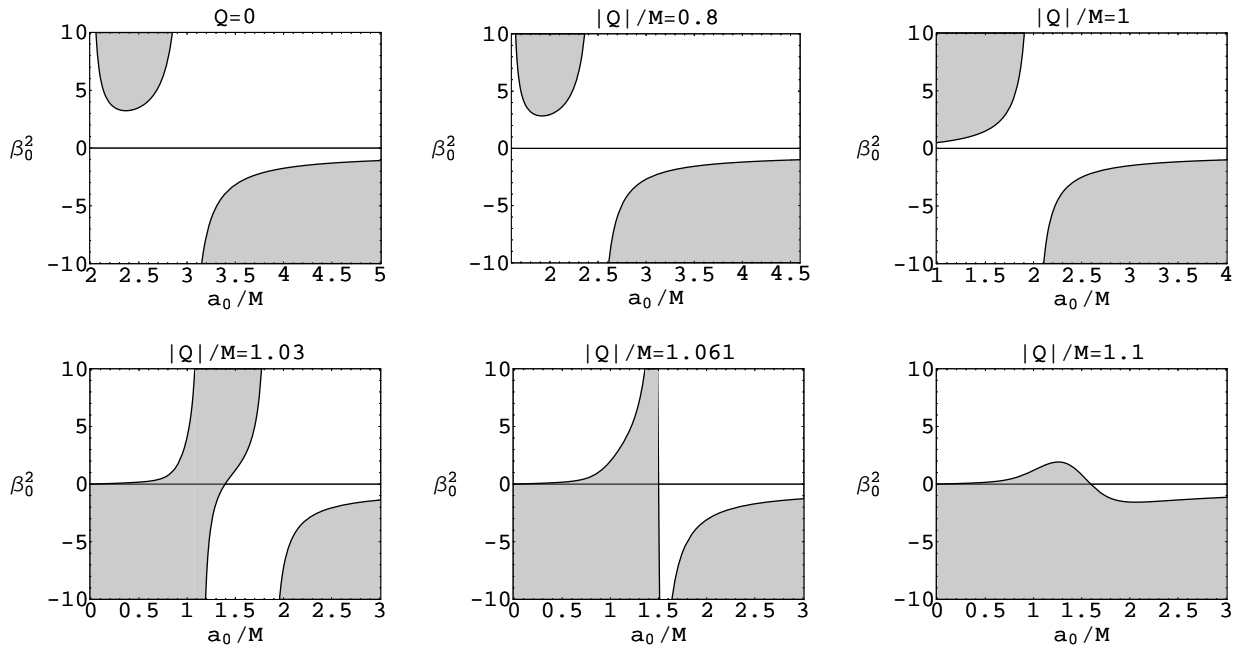}
\end{center}
\caption{Stability regions for Reissner--Nordstr\"om thin--shell wormholes with different values of the charge $Q$. The gray zones correspond to configurations which are stable under radial perturbations. When $Q=0$ the Schwarzschild wormhole is obtained.}
\label{fig1}
\end{figure}
These are the same regions of stability obtained in Ref. \cite{eirom} using the standard method of Sec. \ref{standard} and they are plotted in Fig. \ref{fig1} for different values of charge. A particular case is when $Q=0$, for which Schwarzschild wormholes \cite{poisson} are obtained, with the throat radius that should be greater than the Schwarzschild radius, i.e. $a_{0}>2M$. They are stable when
\begin{equation}
\beta _{0}^{2}>-\frac{a_{0}^2-3Ma_{0}+3M^2}{2(a_{0}-2M)(a_{0}-3M))}\;\;\;\mathrm{if} \;\;\; 2<\frac{a_{0}}{M}<3,
\nonumber
\end{equation}
or
\begin{equation}
\beta _{0}^{2}<-\frac{a_{0}^2-3Ma_{0}+3M^2}{2(a_{0}-2M)(a_{0}-3M))}\;\;\;\mathrm{if} \;\;\;\frac{a_{0}}{M}>3,
\nonumber
\end{equation} 
These stable regions are plotted in Fig. \ref{fig1} (upper panel, left) and they were first found in Ref. \cite{poisson} following the standard approach.\\

The four-dimensional Einstein action with the (scalar) dilaton field $\phi$ coupled to the electromagnetic field $F^{\mu\nu}$, in the Einstein frame, has the form \cite{GHS}
\begin{equation} S=\int d^4x\sqrt{-g}\left(-R+2(\nabla \phi)^2+e^{-2b\phi}F^2\right),
\label{emd}
\end{equation}
where $R$ is the Ricci scalar of spacetime. The parameter $b$ represents the coupling between the dilaton and the Maxwell field \cite{GHS}. When $b=0$, the action corresponds to the usual Einstein--Maxwell scalar theory. For $b=1$, the action was obtained in the context of low energy string theory with a Maxwell field, but with all other gauge fields and antisymmetric field set to zero \cite{GHS}. In the Einstein frame the condition $\delta S=0$ imposed  on the action (\ref{emd}) leads to the Einstein equations with the dilaton and the Maxwell fields as the sources \cite{GHS}:
\begin{equation}
\nabla _{\mu }\left( e^{-2b\phi }F^{\mu \nu }\right) =0,
\end{equation}
\begin{equation}
\nabla ^{2}\phi +\frac{b}{2}e^{-2b\phi }F^{2}=0,
\end{equation}
\begin{equation}
R_{\mu \nu }=2\nabla _{\mu }\phi \nabla _{\nu }\phi +2e^{-2b\phi }
\left( F_{\mu \alpha }F_{\nu }^{\; \alpha }-\frac{1}{4}g_{\mu \nu }F^{2}\right) .
\label{feq}
\end{equation}
These field equations admit spherically symmetric solutions\footnote{There is a Birkhoff theorem if $\phi =\phi (r)$, i.e. $\phi$ does not depend on time \cite{brokome}.} in the form of Eq. (\ref{eq1}), with metric functions, in Schwarzschild coordinates \cite{GHS,mae}, given by
\begin{eqnarray}
f(r) & = &\left( 1-\frac{A}{r}\right)\left( 1-\frac{B}{r}
\right)^{(1-b^2)/(1+b^2)}, \nonumber\\
h(r) & = &r^2\left( 1-\frac{B}{r}\right)^{2b^2/(1+b^2)},
\label{d1}
\end{eqnarray}
where the constants $A,B$ and the parameter $b$ are related with the mass and charge of the black hole by
\begin{eqnarray}
M & =&\frac{A}{2}+\left(\frac{1-b^2}{1+b^2}\right)\frac{B}{2}, \nonumber\\
Q^2&=&\frac{AB}{1+b^2}.
\label{d2}
\end{eqnarray}
In the case of electric charge, the electromagnetic field tensor has non-null components $F_{tr}=-F_{rt}=Q/r^{2}$, and the dilaton field is given by
$\phi =b(1+b^2)^{-1}\ln( 1-B/r)$, where the asymptotic value of the dilaton $\phi_{0}$ was taken as zero. For magnetic charge, the metric is the same, with the electromagnetic field $F_{\theta \varphi}=-F_{\varphi \theta}=Q\sin \theta $ and the dilaton field obtained replacing $\phi $ by $-\phi $. In what follows, we shall consider the generic form of the metric (\ref{d1}), with $0\le b\le 1$. When b=0, the Reissner--Nordstr\"om geometry is obtained. If $b\neq 0$ the metric defined is singular for $r=B$. From Eqs. (\ref{d2}), the constants $A$ and $B$ can be expressed in terms of the mass and the charge:
\begin{equation}
A=M\pm\sqrt{M^2-(1-b^2)Q^2},
\label{d3a}
\end{equation} 
and
\begin{equation}
B=\frac{(1+b^2)Q^2}{M\pm\sqrt{M^2-(1-b^2)Q^2}}.
\label{d3b}
\end{equation}
The plus sign should be taken above if one wants to recover the Schwarzschild metric when $Q=0$ and $b\neq 0$. In this case, for $0\le Q^2 <1+b^2$ the geometry represents a black hole with $B$ and $A$, respectively, the inner and outer horizons; while the outer horizon is a regular event horizon for any value of $b$, the inner one is singular for any $b\neq 0$. If $b\neq 0$ and $1+b^2 \le Q^2 \le 1/(1-b^2)$ we have that $B\ge A$ and the metric corresponds to a naked singularity. For $0<b<1$ the geometry is not well defined if $Q^2 > 1/(1-b^2)$.\\

As it was done in a previous work \cite{dilwh}, from the geometry defined by Eqs. (\ref{d1}) we construct a thin--shell wormhole, assuming that the throat has a radius $a_{0}$ greater than $A$ and $B$ to eliminate the presence of horizons and singularities. Replacing $f(a_{0})$ and $h(a_{0})$ in Eqs. (\ref{eq9a}) and (\ref{eq9b}), we recover the energy density and the pressure at the throat obtained in Ref. \cite{dilwh}
\begin{equation}
\sigma _{0}=-\frac{1}{2\pi a_{0}^2}\left( 1-\frac{A}{a_{0}}\right)^{1/2}\left( 1-\frac{B}{a_{0}}
\right)^{(1-b^2)/(2+2b^2)}\left[ a_{0}+\frac{b^2B}{1+b^2}\left( 1-\frac{B}{a_{0}}
\right)^{-1}\right],
\label{d4}
\end{equation}
\begin{equation}
p_{0}=\frac{1}{8\pi a_{0}^{2}}\left( 1-\frac{A}{a_{0}}\right) ^{1/2}\left( 1-\frac{B}{a_{0}}
\right) ^{(1-b^2)/(2+2b^2)}\left[ 2a_{0}+A\left( 1-\frac{A}{a_{0}}\right) ^{-1}+B
\left( 1-\frac{B}{a_{0}}\right) ^{-1}\right].
\label{d5}
\end{equation}
For the stability analysis, not done previously, we calculate $\Delta $:
\begin{equation}
\Delta =\left( 1-\frac{B}{a_{0}}\right)^{-2b^2/(1+b^2)}
\frac{P_{1}(a_{0})-P_{2}(a_{0})\beta_{0}^2}{4(1+b^2)^2(a_{0}-A) (a_{0}-B)a_{0}^4},
\label{d6}
\end{equation}
where
\begin{eqnarray}
P_{1}(a_{0})&=&(1+b^2)\left\{ -2AB(A+B)+\left[ 3A^2(1+b^2) +6AB+(3+b^2)B^2\right]a_{0}\right.
\nonumber \\
& & \left. -2\left[ 3A(1+b^2)+(3+b^2)B\right]a_{0}^2+4(1+b^2)a_{0}^3\right\}a_{0},
\label{d7a}
\end{eqnarray}
and
\begin{eqnarray}
P_{2}(a_{0})&=&4\left\{ 2A(2+b^2)B^2-B\left[7A(1+b^2)+(3+b^2)B\right]a_{0}\right.
\nonumber \\
& & \left.+(1+b^2)\left[3A(1+b^2)-(-5+b^2)B\right]a_{0}^2-2(1+b^2)^2a_{0}^3)\right\}(a_{0}-A),
\label{d7b}
\end{eqnarray}
are four degree polynomials in $a_{0}$. As we have seen in Sec. \ref{stab}, the static solutions are stable when $\Delta <0$.\\

\begin{figure}[t!]
\begin{center}
\includegraphics[width=17cm]{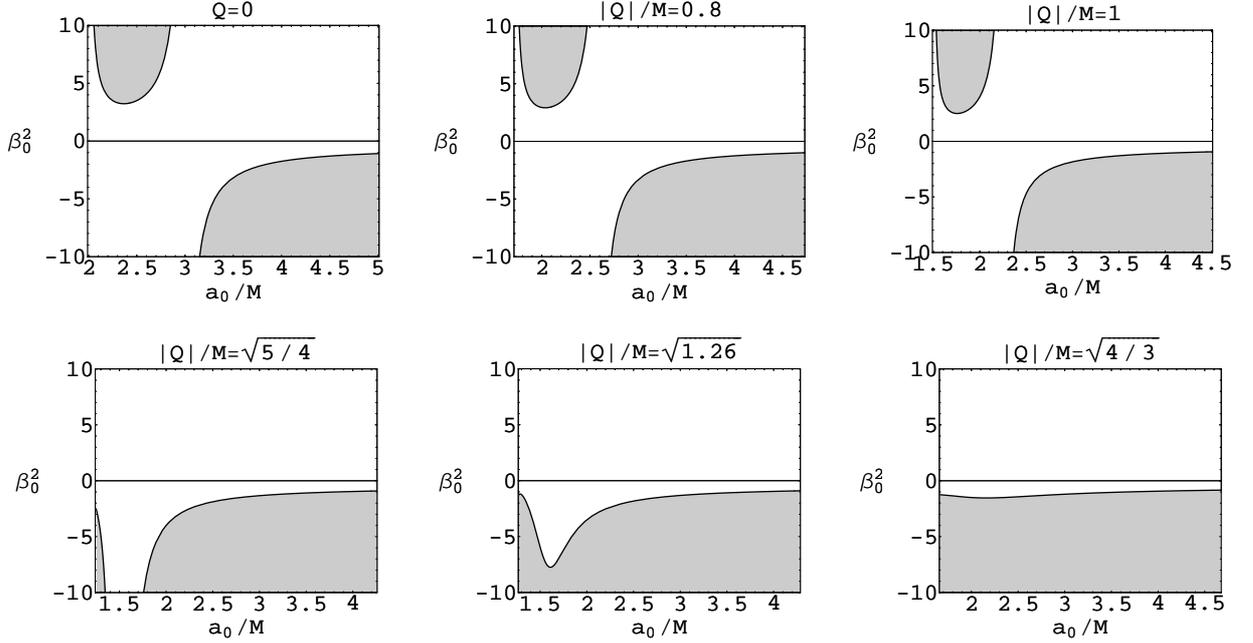}
\end{center}
\caption{Stability regions for dilaton thin--shell spherically symmetric wormholes with $b=0.5$ and different values of the charge $Q$. The gray zones correspond to configurations which are stable under radial perturbations. For $Q=0$ the Schwarzschild wormhole is recovered.}
\label{fig2}
\end{figure}

\begin{figure}[t!]
\begin{center}
\includegraphics[width=17cm]{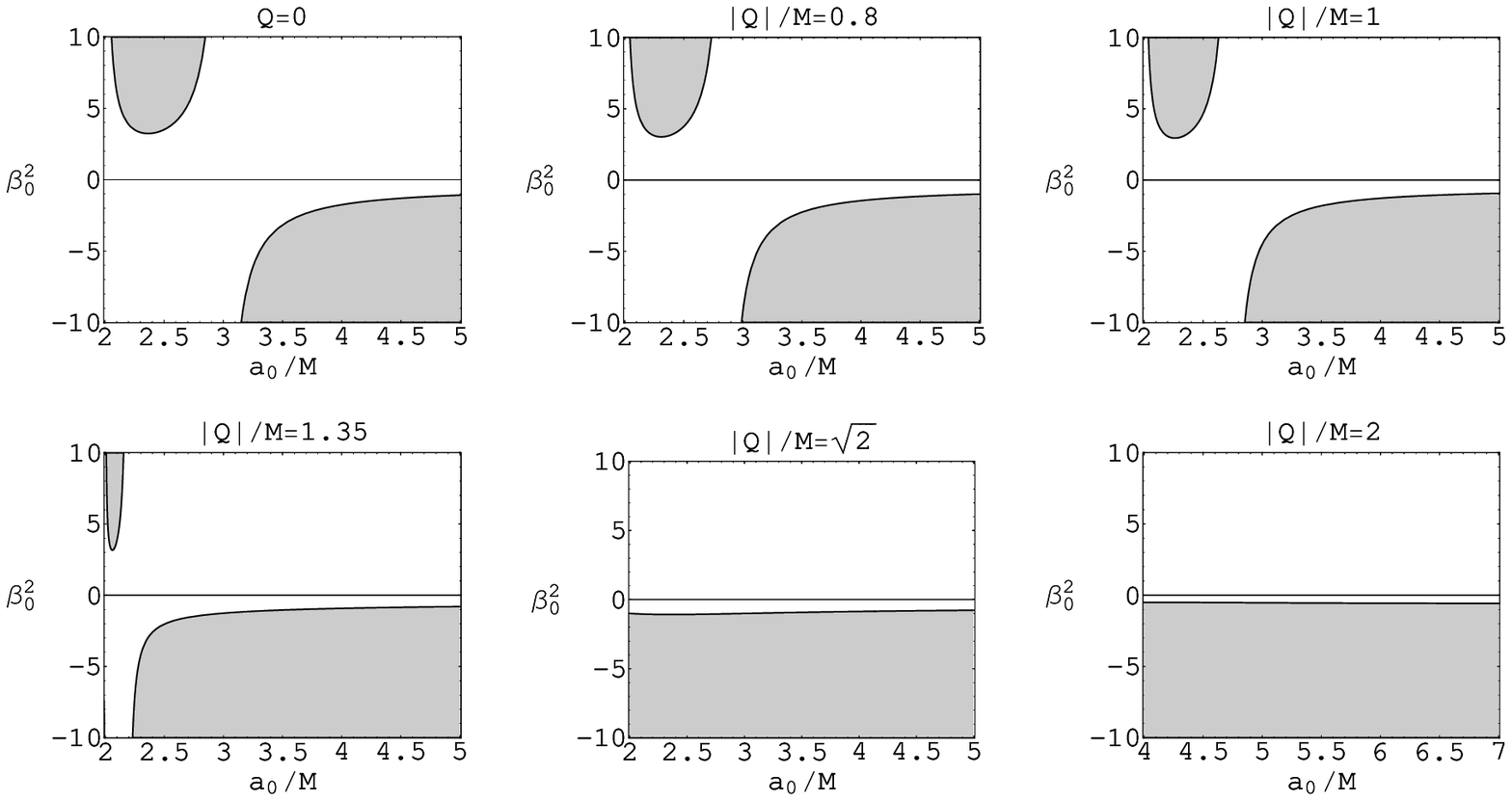}
\end{center}
\caption{Stability regions for dilaton thin--shell spherically symmetric wormholes with $b=1$ and different values of the charge $Q$. The gray zones correspond to configurations which are stable under radial perturbations. For $Q=0$ the Schwarzschild wormhole is recovered.}
\label{fig3}
\end{figure}

If $b=0$ the Reissner--Nordstr\"om wormhole is recovered. When $0<b<1$, the energy density and the pressure at the throat are given by Eqs. (\ref{d4}) and (\ref{d5}), with $A$ and $B$ from Eqs. (\ref{d3a}) and (\ref{d3b}) (plus sign). We take  a throat radius that satisfy $a_{0}>A>B$ if $0\le Q^2<1+b^2$ and $a_{0}>B\ge A$ if $1+b^2\le Q^2\le 1/(1-b^2)$. As pointed above, we cannot take values of charge such as $Q^2>1/(1-b^2)$. From Eq. (\ref{d6}), the stable regions for which the condition $\Delta <0$ is fulfilled correspond to
\begin{equation}
\beta_{0}^2>\frac{P_{1}(a_{0})}{P_{2}(a_{0})}\;\;\; \mathrm{if} \;\;\; P_{2}(a_{0})>0
\nonumber
\end{equation}
or
\begin{equation}
\beta_{0}^2<\frac{P_{1}(a_{0})}{P_{2}(a_{0})}\;\;\; \mathrm{if} \;\;\; P_{2}(a_{0})<0.
\nonumber
\end{equation}
The algebraic complications of four degree polynomials involved makes it cumbersome to write the expressions giving the stability zones explicitly. To see what happens, we choose $b=0.5$ and show in Fig. \ref{fig2} the stability regions for different values of the charge. When $b=1$, the metric functions simplify to give
\begin{equation}
f(r)  = 1-\frac{2M}{r}, \;\;\;\;
h(r)  = r^2\left( 1-\frac{Q^2}{Mr}\right).
\end{equation}
If $0\le Q^2<2M^2$ the geometry corresponds to a black hole with an event horizon in $r_{h}=2M$ and a singular inner horizon in $r_{-}=Q^2/M$, and if $Q^2\ge 2M^2$ represents a metric with a naked singularity in $r=Q^2/M$. The energy density and the pressure at the throat of the wormhole constructed from this metric are given by
\begin{equation}
\sigma _{0}=-\frac{1}{2\pi a_{0}^2}\left( 1-\frac{2M}{a_{0}}\right)^{1/2}\left[ a_{0}+\frac{Q^2}{2M}\left( 1-\frac{Q^2}{a_{0}M}\right)^{-1}\right],
\end{equation}
\begin{equation}
p_{0}=\frac{1}{8\pi a_{0}^{2}}\left( 1-\frac{A}{a_{0}}\right) ^{1/2}\left[ 2a_{0}+2M\left( 1-\frac{2M}{a_{0}}\right) ^{-1}+\frac{Q^2}{M}\left( 1-\frac{Q^2}{a_{0}M}\right) ^{-1}\right],
\end{equation}
where the possible values of the throat radius that eliminate the presence of horizons and singularities are $a_{0}>2M$ if $0\le Q^2<2M^2$ and $a_{0}>Q^2/M$ if $Q^2\ge 2M^2$. In this case, $\Delta $ has the form
\begin{equation}
\Delta =\left( 1-\frac{Q^2}{a_{0}M}\right)^{-1}
\frac{P_{1}(a_{0})-P_{2}(a_{0})\beta_{0}^2}{16(a_{0}-2M) (a_{0}-Q^2/M)a_{0}^4},
\end{equation}
with
\begin{equation}
P_{1}(a_{0})=\frac{8}{M^2}[-2M^3Q^2-MQ^4+(6M^4+3M^2Q^2+Q^4)a_{0}+(-6M^3-2MQ^2)a_{0}^2+2M^2a_{0}^3]a_{0},
\end{equation}
and
\begin{equation}
P_{2}(a_{0})=\frac{16}{M^2}[3MQ^4-Q^2(7M^2+Q^2)a_{0}+2M(3M^2+Q^2)a_{0}^2-2M^2a_{0}^3](a_{0}-2M).
\end{equation} 
The stability regions, where $\Delta <0$, are shown in Fig. \ref{fig3} for relevant values of the charge $Q$. From the figure, we see that the stable regions with $\beta_{0}^2>0$ become smaller as the charge increases. There are no stable configurations with $0<\beta_{0}^2<1$ for any value of charge. By comparing Figs. \ref{fig1}, \ref{fig2} and \ref{fig3} we see that the presence of the dilaton field reduces the possible stable configurations with respect to the Reissner--Nordstr\"om wormholes.

\section{Conclusions}\label{conclu}

In this work a systematic formalism was developed to study the stability under radial perturbations of a general class of spherically symmetric thin--shell wormholes with a linearized equation of state at the throat. The configurations are stable if $\Delta $ is negative, with $\Delta$ a function of the throat radius $a_0$ and the linearization parameter $\beta_0^2$, its functional form depending on the metric adopted to construct the wormholes. The method is straightforward and it was shown equivalent to the standard one, extended in this article to cover all metrics of the form (\ref{eq1}), which is based on the definition of a potential to analyze the stability. The formalism was then applied to the study of dilaton thin--shell wormholes, for which no stability analysis was performed previously. The stable regions in the $(a_{0},\beta _{0}^{2})$ plane are smaller than those corresponding to the Reissner--Nordstr\"om counterparts with the same charge, and they shrink as the coupling parameter $b$ between the dilaton and the Maxwell fields grows.

\section*{Acknowledgments}

The author would like to thank Claudio Simeone for helpful discussions. This work has been supported by Universidad de Buenos Aires and CONICET. Some calculations in this paper were done with the help of the package GRTensorII {\cite{grt}}.

\end{document}